\documentclass[prb,aps,nofootinbib,superscriptaddress,floatfix,10 pt,twocolumn,longbibliography]{revtex4-2}
\usepackage[urlcolor=blue,colorlinks=true,citecolor=blue,linkcolor=blue,bookmarks=false]{hyperref}
\usepackage{amsmath,graphicx,color}
\usepackage{float}
\usepackage{amsfonts}
\usepackage[normalem]{ulem}
\usepackage{orcidlink}
\begin{document}
\title{Origin of the large topological Hall effect in the EuCd$_2$Sb$_2$ antiferromagnet}
\author{Faheem Gul\orcidlink{0000-0002-9705-9900}}
\author{Orest Pavlosiuk\orcidlink{0000-0001-5210-2664}}
\author{Tetiana Romanova\orcidlink{0000-0002-0704-188X}}
\author{Dariusz Kaczorowski\orcidlink{0000-0002-8513-7422}}
\author{Piotr Wiśniewski\orcidlink{0000-0002-6741-2793}}
\email[e-mail: ]{p.wisniewski@intibs.pl}
\affiliation{Institute of Low Temperature and Structure Research, Polish Academy of Sciences, Wrocław, Poland}
%	\pacs{}
\date {\today}
\begin{abstract}
We study the origin of large topological Hall effect in the single-crystalline EuCd$_2$Sb$_2$, which orders antiferromagnetically at the Néel temperature $T_{\rm N}=7.4$\,K. 
Measurements of magnetoresistance and Hall resistivity disclose anomalies that evolve with temperature and magnetic field, closely tracking the magnetization process. 
Analysis of these data identifies three possible mechanisms responsible for the enhanced Berry curvature driving the observed topological Hall effect.
Below and above $T_{\rm N}$, Weyl states are the main sources of large momentum-space Berry curvature, though their formation mechanisms differ in these two temperature ranges. 
Below $T_{\rm N}$, breaking of $C_{3}$ symmetry generates Dirac points that split into Weyl nodes in applied magnetic field, whereas above $T_{\rm N}$, strong spin fluctuations can induce Weyl states. 
The third contribution, which occurs below $T_{\rm N}$, arises from scalar spin chirality developing within antiferromagnetic domain walls, which generates a real-space Berry curvature. 
\end{abstract}
\maketitle
\section{Introduction}\label{Intro}
The discovery of topologically protected electronic states in Dirac and Weyl semimetals prompted extensive research on these systems \cite{Armitage2018rev}.
Electronic bands of these materials are degenerated in specific points of the momentum space. 
That degeneracy is protected by the time-reversal symmetry and/or inversion symmetry \cite{hasan2011three}. %{ando2013topological}
Breaking these symmetries leads to an enhancement of Berry curvature in momentum space, and therefore electrons in applied magnetic field can acquire additional Berry phase, when moving on closed paths. This, in turn, is reflected as intrinsic contributions into the anomalous Hall effect (AHE) \cite{nagaosa2010anomalous,Armitage2018rev}. 
Such contributions are termed "unconventional" to be distinguished from the "conventional" AHE scaling with the spontaneous or field-induced magnetization. The AHE enhanced by unconventional part was reported in several Weyl semimetals, e.g. PrAlGe, Co$_3$Sn$_2$S$_2$ and SmAlSi \cite{meng2019large, Liu2018,Yao2023} and attributed to the rise of Berry curvature in momentum space.

Besides the momentum space, Berry curvature can also emerge in real space due to the presence of spin structures with a finite spin chirality, consequently giving rise to the Berry phase in wavefunctions of conduction electrons \cite{fujishiro2021giant, Taguchi2001, hasan2011three}. 
In both momentum and real space the Berry curvature acts as an apparent magnetic field and manifests itself through magneto-transport anomalies (in both AHE and magnetoresistance) \cite{Xiong2015Science, meng2019large, Pavlosiuk2019, Pavlosiuk2020, thakur2020intrinsic, roychowdhury2023anomalous}.

The unconventional contributions to AHE were reported in various compounds, including Eu-based 1:2:2 antiferromagnetic (AFM) systems \cite{Su2020, Xu2021a, Yan2022, Yi2023, Singh2024}. We will refer to that phenomenon as topological Hall effect (THE), because it is generally attributed to the enhancement of Berry curvature stemming  from various mechanisms.
In the case of hexagonal EuIn$_2$As$_2$ (space group $P6_{\rm 3}/mmc$), a plausible axion insulator, the helical AFM structure gains the spin chirality in applied field, which gives rise to the net Berry curvature in real space \cite{Yan2022}.

In contrast, for compounds EuZn$_2$Sb$_2$, EuZn$_2$As$_2$ and EuCd$_2$As$_2$, crystallizing in a trigonal crystal structure ($P\bar{3}m1$ space group), other mechanisms of THE were proposed, because they are all hosting collinear A-type AFM structures with magnetic moments ferromagnetically aligned in \textit{ab}-planes and stacked antiferromagnetically along the \textit{c}-axes \cite{Singh2024,Wang2022,Soh2019}.

In EuZn$_2$Sb$_2$, it was the spin chirality emerging within the domain walls \cite{Singh2024}, while in EuZn$_2$As$_2$, the local chirality arises (altering the Berry curvature) due to ferromagnetic fluctuations below and above Néel temperature ($T_{\rm N}$) \cite{Yi2023}.
On the other hand, THE exhibited by EuCd$_2$As$_2$ was ascribed to momentum-space Berry curvature from Dirac or Weyl points \cite{Xu2021a}.
All these compounds display significant spin fluctuations due to strong short-range magnetic interactions at temperatures even well above $T_{\rm N}$.
These fluctuations are another source of the pronounced THE persisting above the ordering temperature, as demonstrated by the example of EuCd$_2$As$_2$ \cite{Ma2019}.

However, the small local minimum that appears in topological Hall resistivity, $\rho^{\rm T}_{xy}(H)$, at 2\,K and in magnetic field $\mu_0H\approx 1.3$\,T was overlooked in the analysis presented in Ref.~\cite{Xu2021a}. This anomaly persisted in a form of smaller hump at 6\,K, but disappeared above $T_{\rm N}$. 
This made it different from the main peak of the $\rho^{\rm T}_{xy}(H)$ attaining the largest magnitude at 10\,K, i.e. just above $T_{\rm N}$. We noticed similar small anomalies in other Eu-based 1:2:2 systems that demonstrate THE, described in literature. 
Such fact prompted our research on the origin of THE in a sister material EuCd$_2$Sb$_2$, which was not thoroughly investigated in previous studies \cite{Soh2018, Su2020}.

The compound EuCd$_2$Sb$_2$ (isostructural with EuZn$_2$Sb$_2$, EuZn$_2$As$_2$ and EuCd$_2$As$_2$) has recently been identified as a Weyl semimetal \cite{Soh2018,Su2020}. 
It also hosts A-type AFM structure with spins parallel to the \textit{ab}-plane, \cite{Schellenberg2011,Soh2018} but its $T_{\rm N}$= 7.4\,K is lower than in these three isostructural systems (equal 13.5\,K, 20\,K and 9.5\,K, respectively \cite{Singh2024,Wang2022,Soh2019}).
The rise of AHE in EuCd$_2$Sb$_2$ has previously been attributed to the Berry curvature in momentum space for both single-crystalline  \cite{Su2020}, and thin-film \cite{Nakamura2024a} samples. 
The proposed mechanism was the breaking of $C_{3}$ symmetry in the AFM state leading to the opening of a band gap and emergence of Dirac points in close vicinity to the Fermi level \cite{Soh2018, Su2020}. 
When magnetic field (\textbf{H}) is applied, these Dirac points split into pairs of Weyl points. 
Density functional theory (DFT) calculations reported in Ref.~\cite{Su2020} showed that the field strong enough to induce spin-polarized state (SPS) considerably shifts these points apart. 
The field variation of the AHE data shown in that paper was limited to a single temperature of 2\,K. It displayed a broad minimum in 1\,T and a sharp peak in 2\,T, but this behavior was barely discussed. 
On the other hand, the data collected on thin-film samples displayed completely different behavior with a single maximum (in $\mu_0H\approx2$\,T at $T=2$\,K, and shifting to lower fields with increasing temperature), correlated with anomalies in the magnetoresistance \cite{Nakamura2024a}. 
The resistivity of single-crystalline EuCd$_2$Sb$_2$ has previously been measured in in-plane field ($\mathbf{H}\perp c$) and presented in \cite{Soh2018}, but without specifying if the field was applied transversely or along the current direction. 

This motivated us to carry comprehensive measurements of both magnetoresistance and Hall resistivity on single-crystal samples grown in the same conditions, compare obtained data and bring deeper understanding of the mechanisms of their anomalous behavior.  
Additionally, we measured and analyzed heat capacity of single-crystalline EuCd$_2$Sb$_2$ sample. It allowed the calculation of magnetic entropy, which corroborated the Eu$^{2+}$ ground-state. 
\section {Experimental details}\label{ExpDet}
Single crystals of EuCd$_2$Sb$_2$ were synthesized from Sn-flux (see the Supplemental Material \cite{SupM}).
The stoichiometric composition of the single crystals was confirmed by energy-dispersive X-ray (EDX) spectroscopy using a NanoSEM 230 (FEI Nova) scanning electron microscope equipped with an EDAX Genesis XM4 spectrometer.
Single-crystal X-ray diffraction was performed using a four-circle single-crystal diffractometer (Oxford Diffraction, X'calibur).
The quality and orientation of the single crystals were determined by the Laue backscattering diffraction technique using a LAUE-COS (Proto Manufacturing) system.

Structural characterization of obtained single crystals brought the results, which agree very well  with those reported in earlier works \cite{Schellenberg2011,Soh2018,Su2020}. 
They are presented in the Supplemental Material \cite{SupM}. 

Magnetic properties measurements were performed using a Quantum Design MPMS-XL magnetometer. 
The heat capacity was measured using a relaxation method implemented in a Quantum Design PPMS platform. 
The same PPMS was employed to measure electrical transport using the standard four-point technique on bar-shaped samples cut from the single crystal with a wire saw.
Electrical contacts were made from 50\,$\mu$m-thick silver wire attached to the sample with a silver paint. 
Magnetic-field-dependent longitudinal and Hall resistivities were symmetrized and antisymmetrized, respectively, in order to eliminate spurious contributions from contacts misalignment.
\section{Magnetic Properties}\label{MagProp}

\begin{figure} [b]
	\centering
		\includegraphics[width=\columnwidth]{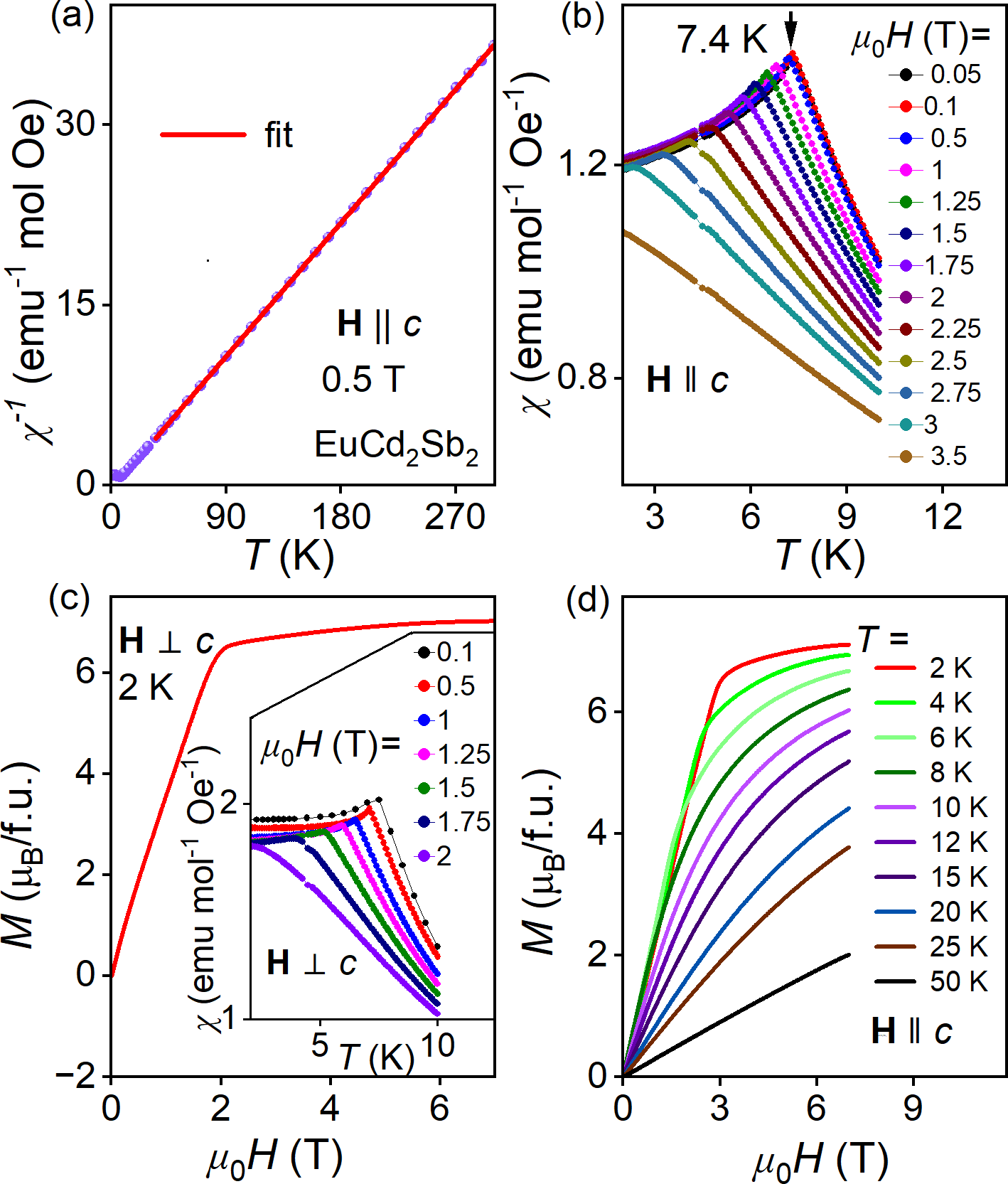}
			\caption{(a) Temperature dependence of $\chi^{-1}$ taken in field $\mu_0H=0.5$\,T, applied along the $c$–axis. Curie-Weiss law fit performed in $T$ range between 40 and 300\,K is represented by straight red line. (b) The $\chi(T)$ measured in various magnetic fields for ${\mathbf H}\parallel c$–axis. (c-d) Magnetization isotherms, $M(H)$: (c) for ${\mathbf H}\perp\!c$ at 2\,K (inset shows $\chi(T)$ measured in various ${\mathbf H}\perp\!c$), and (d) for $\mathbf{H}\parallel c$, at several temperatures.}
				\label{MH-MT}
\end{figure}

Temperature dependence of the inverse magnetic susceptibility, $\chi^{-1}(T)$, magnetic susceptibility, $\chi(T)$, and magnetization as a function of magnetic field, $M(H)$, are shown in Fig.~\ref{MH-MT}.
The $\chi(T)$ measured in a weak magnetic field confirms the AFM ordering at $T_{\rm N}=7.4$\,K, cf. Fig.~\ref{MH-MT}(b), consistent with previous reports \cite{Schellenberg2011,Soh2018,Su2020}. The ordering temperature decreases with increasing magnetic field for its both directions; see Fig.~\ref{MH-MT}(b) and the inset of Fig.~\ref{MH-MT}(c). The Curie-Weiss fit to $\chi^{-1}(T)$ (cf. Fig.~\ref{MH-MT}(a)) gives the paramagnetic Curie temperature $\theta_{\rm P}=3.5$\,K, and effective magnetic moment $\mu_{\rm eff}=8.04\,\mu_{\rm B}$ confirming a divalence of Eu ions (for details see the Supplemental Material \cite{SupM}). 
We note that Soh \textit{et al.} \cite{Soh2018}, most likely by mistake, changed the sign of $\theta_{\rm P}=3.3$\,K derived from their data and also of that cited from \cite{Schellenberg2011} to the negative one. 
Su \textit{et al.} \cite{Su2020} obtained a positive $\theta_{\rm P}=4.6$\,K, in agreement with the work of Schellenberg et al. \cite{Schellenberg2011}.

Because the magnetotransport properties are closely related to the evolution of magnetic structure, we briefly describe that structure here. 
Below $T_{\rm N}$, the $M(H)$ isotherms demonstrate linear behavior in weak magnetic fields, reflecting a tilting of magnetic moments towards the direction of applied field. 
That tilting would lead to the saturation of \textit{M} at the value $M_s=7\mu_{\rm B}/{\rm Eu}$, as expected for Eu$^{2+}$ ions, but the saturation is impeded by thermal fluctuations, and at a certain field $M(H)$ departs from the linearity.

However, when the $M(H)$ is linear, its value is the sum of projections of all magnetic moments on the direction of \textbf{H}. 
For $\mathbf{H}\!\parallel\!{c}$, all moments are tilted from \textit{ab}-planes towards \textit{c}–axis by the same angle $\alpha$, which determines the value of $M=\sin\alpha\,M_s$. 
Extrapolating linear $M(H)$ to the value of $M_s$ yields the magnetic field in which the $M$ would saturate if the process were not disturbed by thermal fluctuations, we denote this field as $H_{\rm s}$. 
In the same way $H_{\rm s}$ values were estimated in several articles on Eu-based 1:2:2 compounds \cite{Soh2018, Soh2019, Pakhira2022, usachov2024magnetism, Wang2022, berry2022type} and used to discuss their magnetic state and anisotropy.
At $T=2$\,K, $H_{\rm s}$ equals 3.1\,T and 1.95\,T, for \textbf{H} applied parallel and perpendicular to the $c$–axis, respectively.
Both values of $H_{\rm s}$ are similar to those reported previously \cite{Soh2018, Su2020}. 
When $T$ increases towards $T_{\rm N}$, the magnetic susceptibility increases, and consequently the $H_{\rm s}$ decreases.
The details of the $H_{\rm s}$ estimation are described in the Supplemental Material \cite{SupM}. 
\begin{figure}[b]
\includegraphics[width=\columnwidth]{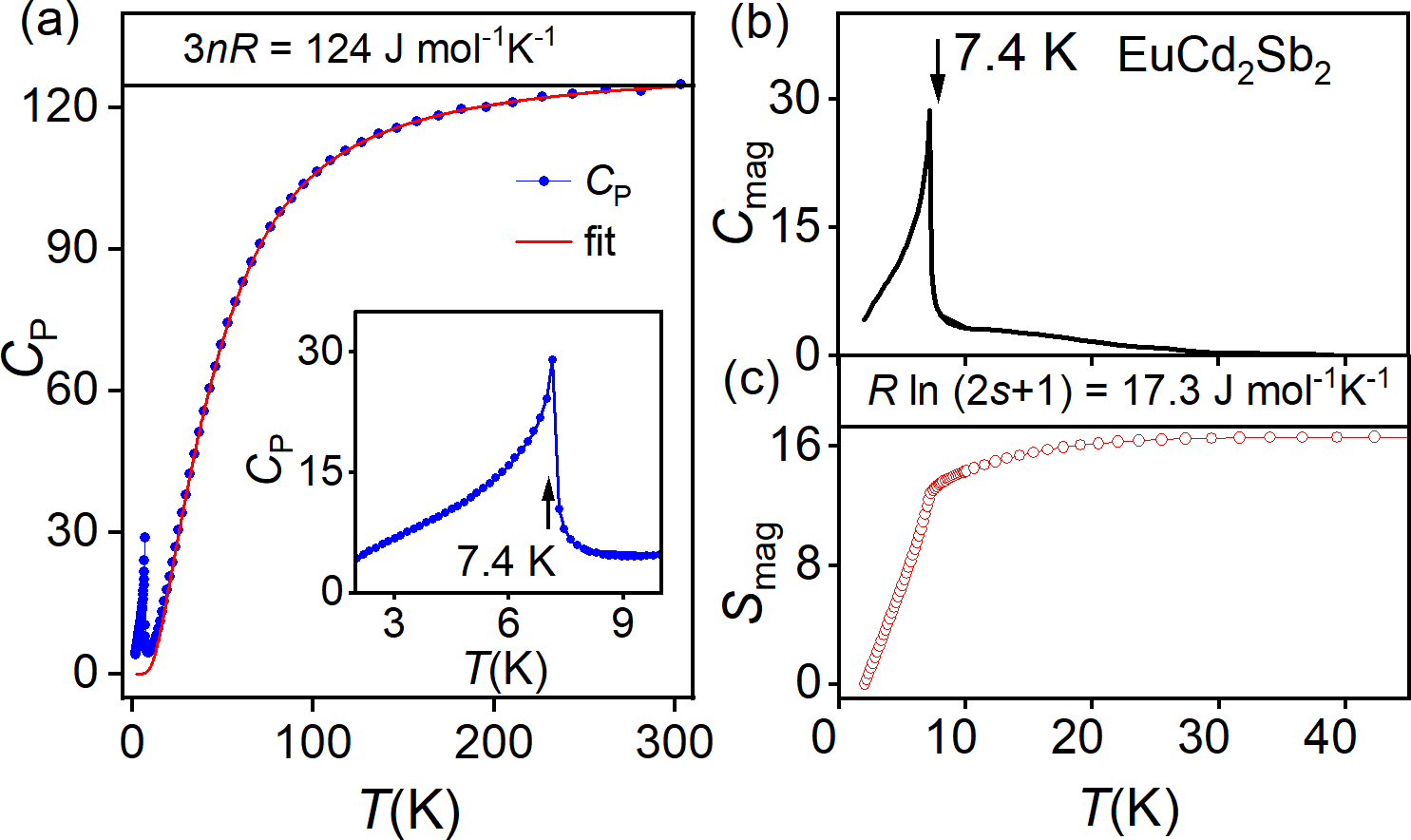}
\caption{(a) Temperature dependence of the heat capacity of EuCd$_2$Sb$_2$. The red solid line shows the fit of Eq.~1 %\ref{eq_HeatC}
to $C_p(T)$. The inset shows the heat capacity anomaly due to the phase transition to AFM ordered state. (b) Temperature dependence of the magnetic contribution $C_\text{mag}$ to heat capacity, and (c) the magnetic entropy ($S_{\rm mag}$). The $C_p, C_{\rm mag}$ and $S_{\rm mag}$ are all shown in $\rm J\,mol^{-1} K^{-1}$ units.} \label{HCF}
\end{figure}
\section{Heat capacity}\label{HeatCap}
The heat capacity ($C_p$) of EuCd$_2$Sb$_2$ has hitherto been measured only on polycrystalline samples, and only a value at room temperature and the anomaly at 7.13\,K have been described \cite{Zhang2010}. 
We remeasured it on a single crystal and performed detailed analysis. 
The temperature dependence of the heat capacity is shown in Fig.~\ref{HCF}(a).
A $\lambda$-shaped peak is observed at low temperatures, as shown in the inset of Fig.~\ref{HCF}(a). 
This corroborates $T_{\rm N}$ value of 7.4\,K, in accordance with the results of all previous magnetic and electrical transport property measurements on single crystals of the compound.
The maximum value of $C_{\rm p}(T)$ %126\,J\,mol$^{-1}$K$^{-1}$
 obtained at $T=300$\,K is nearly identical to the classical Dulong-Petit limit of $3nR=124.71\,{\rm J}\,{\rm mol}^{-1}$K$^{-1}$, where $n$ is the number of atoms per unit cell and $R$ is the universal gas constant. 
To calculate magnetic entropy, we assumed that the total heat capacity can be described by the formula
\begin{equation}
C_{\rm p}=\gamma T+ p\,C_{\rm E}+ (1-p)\,C_{\rm D}+C_{\rm mag}, 
\label{eq_HeatC}\end{equation}
and fitted it to our data for the $40\,{\rm{K}}\!<T\!<\!300$\,K range (with magnetic contribution, $C_{\rm {mag}}$, assumed negligible due to the absence of critical spin fluctuations, and fixed at zero value).
Here $\gamma T$ is the electronic contribution, $\gamma$ is the Sommerfeld coefficient, 
and $p$ denotes the relative weight of the Einstein phonon term.
Einstein term 
$C_{\rm E}=3nR \left(\frac{\theta_{\rm E}}{T}\right)^2 \frac{e^{\theta_{\rm E}/T}}{(e^{\theta_{\rm E}/T}-1)^2}$ 
and Debye term\\ $C_{\rm D}$=$9nR\left(\frac{T}{\theta_{\rm D}}\right)^3\int_0^{{\theta_{\rm D}}/{T}}\frac{x^{4} e^{x}}{(e^{x}-1)^{2}}dx,$ 
describe contributions from optical and acoustic phonon modes respectively. $\theta_{\rm E}$ and $\theta_{\rm D}$ are Einstein and Debye temperatures, respectively. 
The fit is shown in Fig.~\ref{HCF}(a) and the obtained parameters are $\gamma=8.1$\,mJ\,mol$^{-1}$K$^{-2}$, $\theta_{\rm E}=81$\,K, $\theta_{\rm D}=228$\,K, and $p=0.36$, indicating that acoustic modes are dominant in the lattice specific heat. These characteristic temperatures are close to that for EuZn$_2$Sb$_2$ \cite{Singh2024} and lower than that for EuMg$_2$Sb$_2$\cite{Pakhira2022}, (cf. Table S2 in the Supplemental Material \cite{SupM}). 
The $C_{\rm mag}$ contribution was determined for $2\,{\rm{K}}\!<T\!<\!40$\,K range by subtracting from the measured data the nonmagnetic contribution $(\gamma T+ p\,C_{\rm E}+ (1-p)\,C_{\rm D})$
calculated using the parameters obtained from the fit described above. 
Magnetic entropy was calculated using the relation $S_{\rm mag}(T)=\int_{2 {\rm K}}^{T} (C_{\rm mag}(x)/x)\,dx$.
At $T=35$\,K, it saturates at 16.6\,J\,mol$^{-1}$K$^{-1}$ (cf. Fig.~\ref{HCF}(c)), fairly close to the theoretical value for the divalent europium ion $R\,{\rm ln}(2S+1)=17.3$\,J\,mol$^{-1}$K$^{-1}$ (with $S=$7/2 the spin of Eu$^{2+}$).
We estimated the density of states ($N$) at the Fermi level, using the obtained $\gamma$ value and the formula $N(E_{\rm F})= {3\gamma }/{\pi^{2}N_{\rm A}k_{\rm B}^{2}}$, where $k_{\rm B}$ and $N_{\rm A}$ are the Boltzmann and Avogadro constants, respectively. 
The obtained value, $N(E_{\rm F})=3.6$\,states/eV lies in between the theoretically predicted values of 1\,states/eV \cite{Soh2018} and 13\,states/eV \cite{Zhang2010}. Note that the former theoretical value was calculated in conjunction with the experimental angle-resolved photoemission spectroscopy (ARPES) results. 
	\begin{figure}[h]
		\centering
		\includegraphics[width=\columnwidth]{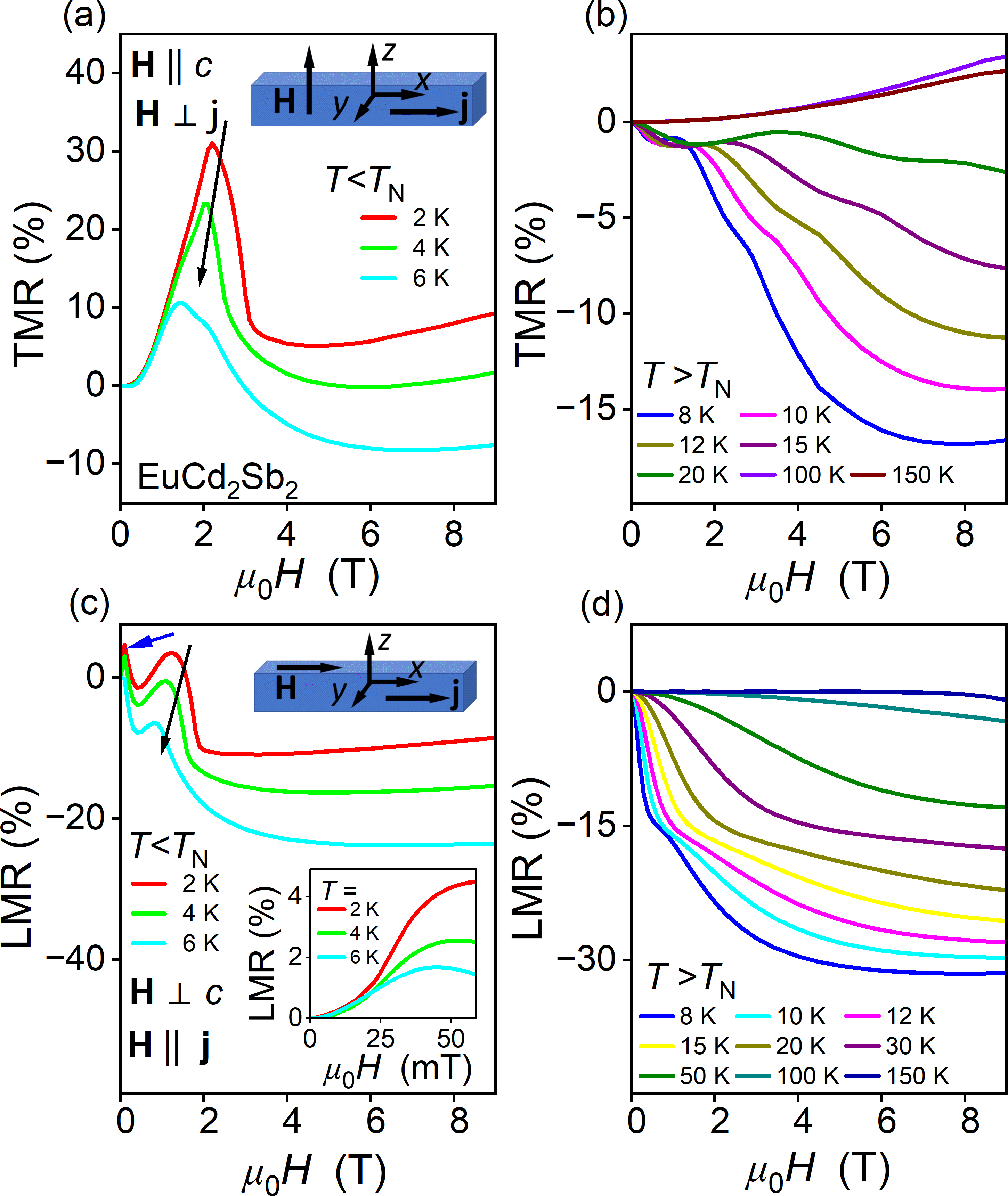}
		\caption{Transverse (a, b) and longitudinal (c,d) magnetoresistance of the EuCd$_2$Sb$_2$ as a function of magnetic field for different temperatures ranging from 2 to 150\,K. (a) and (c) show data collected at $T<T_{\rm N}$, (b) and (d) those collected at $T>T_{\rm N}$. Inset to (c) shows a blowup of the low-field region. The measurement geometries are shown in the insets (with \textit{z} parallel to crystallographic \textit{c}–axis).}
		\label{TMR_LMR}
	\end{figure}
\section{Magnetoresistance}\label{secMR}
In Fig.~\ref{TMR_LMR}, we present field variation of the magnetoresistance (MR) measured in both transverse (in $\mathbf{H}$ applied perpendicular to the direction of electrical current $\mathbf{j}$, denoted as TMR) and longitudinal  (in $\mathbf{H}\!\parallel\mathbf{j}$, denoted as LMR) geometry at different temperatures. 
Electrical current was always flowing in \textit{ab}-plane, and MR values were calculated as $[(\rho(H)/\rho(H=0))-1]$.
For both TMR and LMR, the distinct peaks are evident, marked with black arrows in Figs.~\ref{TMR_LMR}(a) and \ref{TMR_LMR}(c). 
A similar peak has previously been observed for EuCd$_2$Sb$_2$ in TMR at 2\,K, but only a weak hump appeared in LMR, as reported in Ref.~\cite{Su2020}. 
Below $T_{\rm N}$, in both geometries, MR attains maxima in magnetic fields close to $H_{\rm s}/\sqrt{2}$. 
Their magnitudes decrease and maxima shift to the lower magnetic field with increasing temperature. 
For example, the rise in TMR is nearly 31\% and 4.9\% in LMR at 2\,K, whereas at $T=4$\,K the corresponding values are 23\% and 3.4\%, respectively (see Fig.~\ref{TMR_LMR}(a) and (c)).
In the high magnetic field regime, a slight upturn in both TMR(\textit{H}) and LMR(\textit{H}) is observed at temperatures below $T_{\rm N}$ (see Figs.~\ref{TMR_LMR}(a) and (c)). 
Large negative TMR and LMR are observed at 8 K, with values of $-17$\% and $-31.3$\%, respectively. 
Above 8\,K, both negative LMR and negative TMR start to diminish with increasing temperature, with TMR becoming positive at $T\ge$ 100\,K, while LMR becomes very small ($\approx-1$\%) at 150\,K as shown in Figs.~\ref{TMR_LMR}(b) and (d), respectively.
The enhanced negative LMR in EuCd$_2$Sb$_2$ has been tentatively ascribed to the emergence of the Weyl points in the applied magnetic field and the resulting chiral magnetic anomaly \cite{Su2020}. 
Below $T_{\rm N}$, additional positive contribution is observed in the low-field region of LMR(\textit{H}) shown in inset to Fig.~\ref{TMR_LMR}(c), which might be due to the weak antilocalization. 

Comparing our LMR(\textit{H}) to the resistance of single-crystalline EuCd$_2$Sb$_2$ measured in in-plane field ($\mathbf{H}\perp c$) by Soh \textit{et al.} \cite{Soh2018}, one notice that their MR values in $\mu_0H=5$\,T (spanned between -4.6\% and -17\%, at $T=2$\, and 8\,K, respectively), are close to our TMR. On the other hand, the field-positions of maxima in their MR are very similar to those in our LMR (close to $H_s/\sqrt{2}$, with $H_s=1.95$\,T with $\mathbf{H}\perp c$). This indicates that the data in Ref.~\cite{Soh2018} were  collected in $\mathbf{H}\perp\,\mathbf{j}\perp c$ and $\mathbf{H}\perp c$ (i.e., in TMR geometry), and that the field-positions of MR anomalies depend on the spin structure, but not on the current direction within \textit{ab}-plane (which most likely was different in our measurements and that of ref.~\cite{Soh2018} ). 

The overall picture of MR in the semimetallic compounds listed in Table \ref{parameters-table1} is in sharp contrast to that reported for isotypic but semiconducting EuCd$_2$P$_2$ and EuZn$_2$P$_2$, displaying colossal negative magnetoresistance ascribed to magnetic polarons, whose occurrence is enabled by much smaller carrier concentrations ($10^{16}-10^{17}\mathrm{cm}^{-3}$) \cite{Li2024,Cook2025}. 
\begin{table}[]
\begin{center}
\caption{Characteristic features of selected Eu-based 1:2:2 compounds which show pronounced AHE well above $T_{\rm N}$ due to different mechanisms (see text for details).}
\label{parameters-table1}
\begin{tabular}{c c c c c }
 \hline
  & $T_{\rm N}$\,(K) & $\Delta \rho_{sf}$ & $n_{\rm H}$\,(cm$^{-3}$)
  & Ref.\\ [0.6ex]
 \hline
 \hline
  EuIn$_2$As$_2$ & 17  & 0.42 & 4.1$\times10^{19}$ & \cite{Yan2022} \\
 \hline
  EuZn$_2$As$_2$ & 20  & 23 & 4.1$\times10^{18}$ &\cite{Yi2023}  \\
 \hline
  EuZn$_2$Sb$_2$ & 13.5 & 0.016 & 2.2$\times10^{19}$ &\cite{Singh2024} \\
  \hline
  EuCd$_2$As$_2$ & 9.5 & 1 & 8.0$\times10^{19}$ &\cite{Xu2021a}\\
 \hline
 EuCd$_2$Sb$_2$ & 7.4  & 0.044 & 2.3$\times10^{19}$& this work  \\
 \hline
\end{tabular}
\end{center}
\end{table}

%\begin{widetext}

\begin{figure}[h]
	\centering
	\includegraphics[width=\columnwidth]{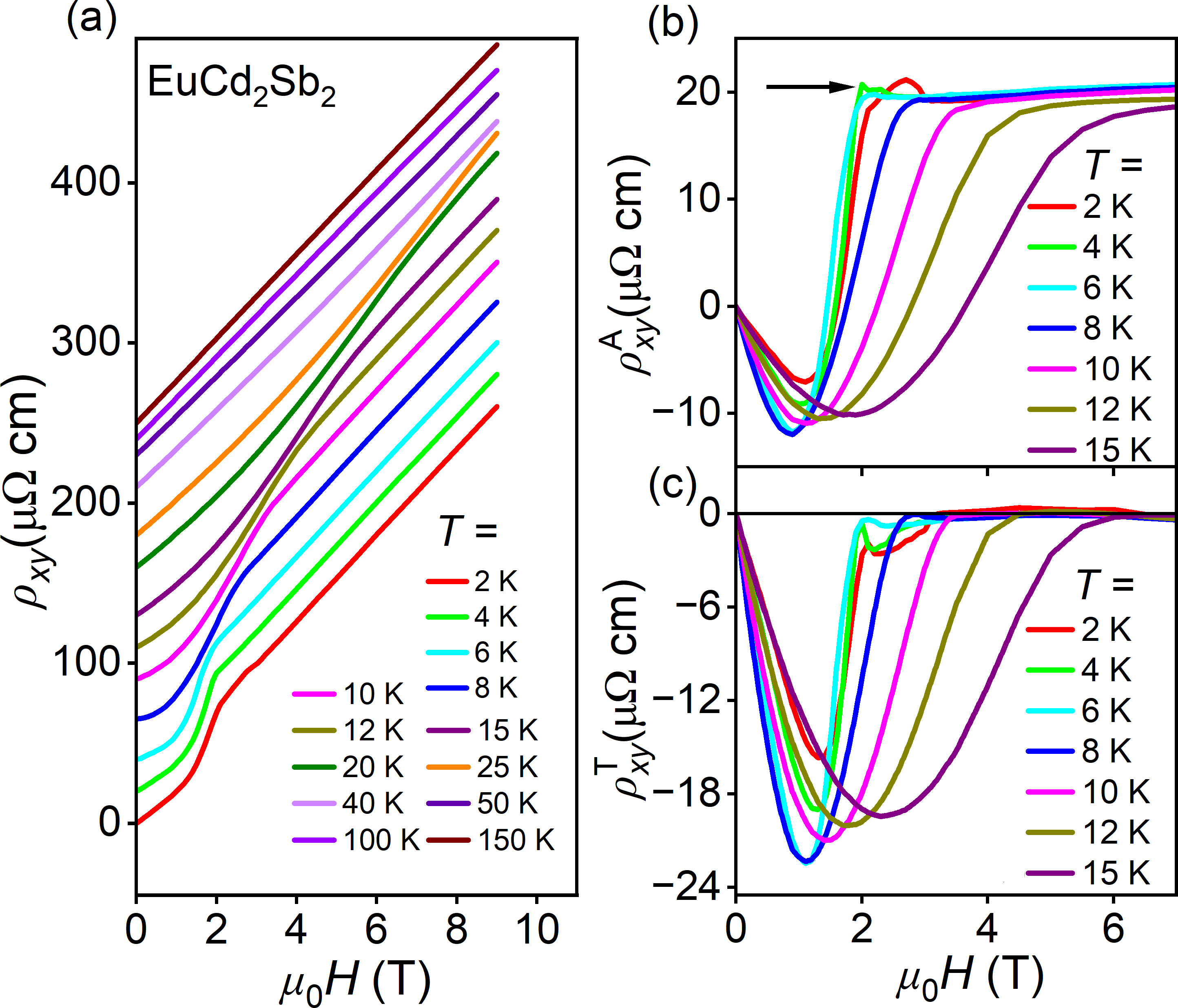}
	\caption{Magnetic field dependence of the Hall resistivity of EuCd$_2$Sb$_2$. (a) $\rho_{xy}(H)$ for various temperatures in the range 2–150\,K (the lines for $T\geq4$\,K are offset for clarity). (b) Anomalous Hall resistivity $\rho_{xy}^{\rm A}(H)$ at several temperatures, an arrow marks the region in which there are additional humps at low temperatures. (c) Topological Hall resistivity as a function of the magnetic field at several temperatures.}
	\label{TotalHall}
\end{figure}
%\end{widetext}
\section{Hall resistivity}\label{HALLresist}
Fig.~\ref{TotalHall}(a) shows the magnetic field dependence of the Hall resistivity, $\rho_{xy}(H)$, at various temperatures.
At $T\geq50$\,K, $\rho_{xy}$ is linear in $H$, whereas at lower temperatures a curvature is clearly visible, indicating the emergence of anomalous Hall effect. 
The slopes of $\rho_{xy}(H)$ are almost identical for $T\geq50$\,K, and also for $T\leq15$\,K, but in the latter case only at fields above certain values, which are very close to $H_{\rm s}$ when $T<T_{\rm N}$. 
Also very similar are the slopes just above zero field at 2 and 4\,K, where the magnetic structure is robust and fluctuations are weakest. Temperature-independent band structure of EuCd$_2$Sb$_2$ was confirmed by ARPES measurements in the 15–121\,K range (cf. supplementary Fig.~S2 in Ref.~\cite{Ma2019}).
All these justify the assumption that the ordinary Hall resistivity is rectilinear in $H$ (with the slope very slightly changing with $T$ and $H$), while the curvature of $\rho_{xy}(H)$ in magnetic fields below $H_{\rm s}$ is due to the anomalous Hall effect. 

Both the Hall conductivity, $|\sigma_{xy}|= 400\,(\Omega\rm{cm})^{-1}$ and the longitudinal conductivity $\sigma_{xx}= 1.7\times10^3\,(\Omega\rm{cm})^{-1}$ (calculated as shown in the Supplemental Material \cite{SupM}) place EuCd$_2$Sb$_2$ in the region of the scaling diagram proposed in \cite{nagaosa2010anomalous} where the skew-scattering contribution is negligible. 
 Hence, by ruling out that contribution, the Hall resistivity for a single band system can be described as a sum of three contributions \cite{nagaosa2010anomalous}:
	\begin{equation}\label{Eq1}
		\rho_{xy}=\rho_{xy}^{\rm o}+\rho_{xy}^{\rm CA}+ \rho_{xy}^{\rm T}%=  R_{ o}\mu_0H+{S_{\rm H }}\rho_{xx}^{2} M + \rho_{xy}^{\rm T}
	\end{equation}
where $\rho_{xy}^{\rm o}=R_{\rm H}H$ is the ordinary Hall resistivity, $\rho_{xy}^{\rm CA}=S_{\rm H}\rho_{xx}^{2}M$ is the conventional anomalous Hall resistivity, and $\rho_{xy}^{\rm T}$ is topological Hall effect contribution. 
Here $S_{\rm H}$ and $R_{\rm H}$ are anomalous Hall and ordinary Hall coefficients, respectively. 
The slope of the magnetic field dependence of the ordinary Hall contribution, $R_{\mathrm H}H$, defines the carrier concentration $n_{\rm H}=1/eR_{\rm H}$. 

The observed positive ordinary Hall coefficient indicates that holes are the majority charge carriers. 
Given the fact that the $M(H)$ tends to saturate in magnetic field of about 3.1\,T for $\mathbf{H}\parallel\,c$–axis, it is reasonable to use the slope of linear $\rho_{xy}(H)$ from the high magnetic field region to calculate $R_{\rm H}$. The carrier concentration at different temperatures is provided in Table~\ref{regions}.
The carrier concentration in EuCd$_2$Sb$_2$ is very modestly affected by temperature. At $T=2$\,K, $n_{\rm H}=2.34\times10^{19}\text{cm}^{-3}$. It is larger by only 5\% at $T=100$\,K, indicating that the Fermi surface almost does not change with temperature.   
Previously reported $n_{\rm H}$ for EuCd$_2$Sb$_2$ at 2\,K was $2.5\times10^{19}$\,cm$^{-3}$ \cite{Su2020}, which is very close to the value we obtained, indicating similar electronic structure of samples.  

Next, to find the $S_{\rm H}$, we used the relation; $\rho_{xy}^{\rm A}= {S_{\rm H }}\rho_{xx}^{2}M+\rho_{xy}^{\rm T}$, and assumed that  $\rho_{xy}^{\rm T}$ vanishes in  magnetic field above $H_{s}$. This assumption stands on the basis that the slope of the $\rho_{\rm xy}(H)$ is almost temperature independent in a high magnetic field range, so there $\rho_{xy}^{\rm A} = {S_{\rm H }}\rho_{xx}^{2} M$. The value of  $\rho_{xy}^{\rm A}$ in the highest magnetic field was used to estimate $S_{\rm H}$. Then the $M(H)$ and $\rho_{xx}(H)$ data were used to calculate $\rho_{xy}^{\rm A}$ for the whole field range. Steps of the extraction of $\rho_{\rm xy}^{\rm T}(H)$ for $T$ = 2\,K  are shown in Fig.~S3(a) in the Supplemental Material \cite{SupM}. The same $S_{\rm H}$ was used to subtract $\rho_{\rm xy}^{\rm CA}$ from $\rho_{\rm xy}^{\rm A}$ at higher temperatures. 

\begin{table}%[!h]
%\begin{center}
\caption{Ranges of magnetic field used in fitting to estimate the ordinary Hall resistivity for various temperatures up to 150\,K, and the respective carrier concentration $n_{\rm H}$ obtained from the slopes of straight lines fitted to $\rho_{xy}(H)$ in these field ranges.}
 \label{regions}
 \begin{center}
\begin{tabular}{r c c}
%\hline
\textit{T} (K)  &~~~~$\mu_0$\textit{H}\,(T) ~~~~& $n_{\rm H}\,(\rm{cm}^{-3})$  \\ [0.5ex]
 \hline\hline
2–15 & 7 – 9 & 2.3$\times 10^{19}$\\
  50 & 0 – 9 & 2.5$\times 10^{19}$\\
 100 & 0 – 9 & 2.4$\times 10^{19}$\\
 150 & 0 – 9 & 2.4$\times 10^{19}$\\
 \hline
\end{tabular}

\end{center}
\end{table}

The topological Hall resistivity, $\rho_{xy}^{\rm T}(H)$, is due to the real- or momentum-space Berry phase stemming from a continuous change of the spin configuration in applied field. We assume that it vanishes in the high-field region, where the spin-polarized state (SPS) is induced. This is obvious for the real-space case, but the momentum-space Berry phase depends on Zeeman band splitting, which induces Weyl nodes and modulates their mutual separation and distance from the Fermi level. Therefore, in some ranges of magnetic field, a corresponding part of $\rho_{xy}^{\rm T}(H)$ might mirror the behavior of magnetization, and therefore be difficult to separate from $\rho_{xy}^{\rm CA}(H)$. 
However, the shape of the prominent anomalies in $\rho_{xy}^{\rm A}(H)$ (obtained by subtraction of the linear-in-field term from the $\rho_{xy}(H)$ and shown in Fig.~\ref{TotalHall}(b)) and their convergence at $\approx 20\,\mu\Omega {\rm cm}$ for all $T\leq$15\,K, indicates that this part of $\rho_{xy}^{\rm T}(H)$ becomes negligible in fields close to $H_{\rm s}$.  
Subsequently, the topological Hall resistivity, $\rho_{xy}^{\rm T}$, was extracted using Eq.~\ref{Eq1}, and it is shown in Fig.~\ref{TotalHall}(c).

The origin of unconventional AHE in ferromagnets is often assessed from the dependence of the corresponding conductivity, $\sigma_{xy}^{\rm T}$, on the longitudinal conductivity, $\sigma_{xx}$.\\ According to a scaling scheme proposed in Ref.~\cite{nagaosa2010anomalous}, the Berry phase is responsible for AHE when $\sigma_{xy}^{\rm A}$ is independent of $\sigma_{xx}$, in a regime of $[10^4\,(\Omega{\,\rm cm})^{-1} < \sigma_{xx} < 10^6\,(\Omega{\,\rm cm})^{-1}]$, whereas the extrinsic side jump mechanism is the source of AHE, if  $\sigma_{xy}^{\rm A}\propto{\sigma_{xx}}^\alpha$, with $\alpha\approx$1.6, for $[\sigma_{xx} < 10^4\,(\Omega {\,\rm cm})^{-1}]$. 

\begin{figure} [h]
%\centering
\includegraphics[width=\columnwidth]{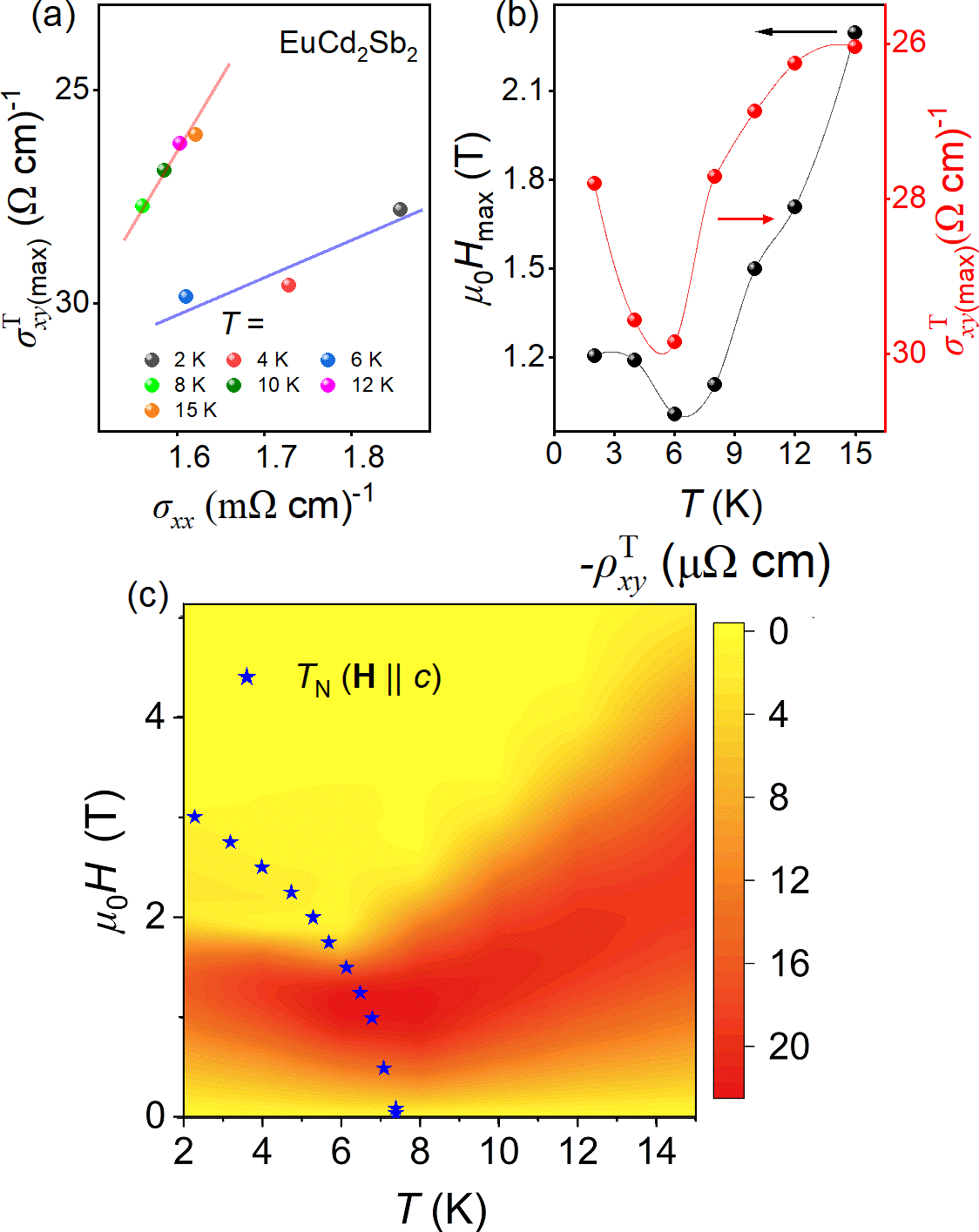}
\caption{ (a) The extrema of topological Hall conductivity, $\sigma_{xy}^{\rm T}$ versus the longitudinal conductivity, $\sigma_{xx}$. (b) Temperature dependence of these extrema of $\sigma_{xy}^{\rm T}$ (right axis) and temperature dependence of the magnetic field strength, in which they occur, $\mu_0 H_{\text{min}}$, (left axis) - lines are guides to the eye. (c) The color map of the topological Hall resistivity, $\rho_{xy}^{\rm T}$ plotted in ($T$,$H$) coordinates superimposed on the phase diagram for $\mathbf{H}\parallel c$, constructed based on the magnetic measurements. Stars mark $T_{\rm N}$ values in different fields, derived from the Supplemental $M(T)$ data \cite{SupM}.} 
\label{origin-signature}
\end{figure}

In order to test that scaling on our data we calculated the respective conductivities as: 
$\sigma_{xx}=\rho_{xx}/(\rho^{2}_{xx}+\rho^{2}_{xy})$ and 
$\sigma^{\rm T}_{xy}=-\rho_{xy}^{\rm T}/(\rho^{2}_{xx}+\rho^{2}_{xy})$ (cf. the Supplemental Material \cite{SupM}).
Then, the maximal values of $\sigma_{xy}^{\rm T}(H)$ (shown in supplemental Fig.~S5(c) \cite{SupM}) were plotted against $\sigma_{xx}$ at various temperatures up to 15\,K, which is shown in Fig.~\ref{origin-signature}(a).

EuCd$_2$Sb$_2$ is a moderately bad metal, with $\sigma_{xx}\approx 1.7\times 10^3\,\Omega{\,\rm cm}$ placing it at the border between regimes of extrinsic and intrinsic AHE, proposed for ferromagnets within the above-mentioned scaling scheme. Actually, its $\sigma_{xy}^{\rm A}$ changes weakly with  $\sigma_{xx}$ above $T_{\rm N}$, and even weaker at $T<T_{\rm N}$, when compared to ferromagnets in the extrinsic-AHE regime \cite{fujishiro2021giant}. A small number of experimental point makes virtually impossible to precisely estimate the scaling exponent in $\sigma_{xy}^{\rm T}\propto{\sigma_{xx}}^\alpha$, but Fig.~\ref{origin-signature}(a) shows that the $\alpha\approx1$ approximation is reasonable. 
Of course that scheme must be adopted to AFM systems with caution. Serious departures from that scaling was demonstrated for numerous AFM systems, e.g. MnGe and SmAlSi \cite{fujishiro2021giant,Gao2025}  
Both the magnitudes of conductivities and their scaling in EuCd$_2$Sb$_2$ strongly resemble those shown for GdPtBi \cite{Suzuki2016}, thus we believe that they hint at the predominantly intrinsic nature of THE.  
Nevertheless, we observe that $\sigma_{xy}^{\rm T}$ is dependent on $\sigma_{xx}$, and two distinct approximately linear regimes can be identified, with clearly different slopes below and above $T_{\rm N}$ (shown by red and blue lines in Fig.~\ref{origin-signature}(a)). 
Figure~\ref{origin-signature}(b) shows the temperature dependence of both $H_{\rm max}$, the magnetic field at which $\sigma_{xy}^{\rm T}$ achieves its maximum value, and that value $\sigma_{xy(\rm max)}^{\rm T}$ itself.
Both quantities decrease with increasing $T$ below $T_{\rm N}$ and increase above $T_{\rm N}$. 
This indicates that distinct mechanisms are responsible for THE-related dips in these two temperature regions. 
The Hall angle $\theta_{\rm H}$, calculated as $\sigma_{xy}^{\rm T}/\sigma_{xx}$, attains a maximum of 2.6\% for 6\,K (cf. Fig. S3(b) in the Supplemental Material \cite{SupM}).
In topological semimetals, where $\theta_{\rm H}$ originates purely from momentum-space Berry curvature, the values of $\theta_{\rm H}$ are an order of magnitude larger.
For example, in the well-established Weyl semimetals GdPtBi and TbPtBi, $\theta_{\rm H}$ exceeds 20\% \cite{Suzuki2016, Shekhar2018, Pavlosiuk2020}. 
In contrast, the maximum Hall angle associated with the THE is about 1\% in both EuCd$_2$As$_2$ and EuZn$_2$Sb$_2$ \cite{Xu2021a, Singh2024}, close to that we obtained for EuCd$_2$Sb$_2$.

Below $T_{\rm N}$, given the A-type AFM structure of our system, we can safely exclude the mechanism of the THE proposed for EuIn$_2$As$_2$, which arises from its helical magnetic structure \cite{Yan2022}.
On the other hand, for EuZn$_2$As$_2$ (also A-type AFM), the width of the peak in $\rho^{\rm A}_{xy}(H)$ associated with the THE decreases with increasing temperature in both the paramagnetic and antiferromagnetic phases, while the field position of the extremum ($\approx$0.39\,T) remains nearly temperature independent \cite{Yi2023}.
In contrast to EuZn$_2$As$_2$, both EuCd$_2$As$_2$ \cite{Xu2021a} and EuCd$_2$Sb$_2$ (Fig.~\ref{TotalHall}(d)) demonstrate the opposite trend in the region above $T_{\rm N}$, where the THE-related anomalies broaden significantly with increasing temperature, and the field positions of their extrema show a pronounced temperature dependence, both below and above $T_{\rm N}$.

We suggest that the THE, manifesting as broad dips of $\rho_{xy}^{\rm T}(H)$, both below and above $T_{\rm N}$, can be primarily attributed to momentum-space Berry curvature, which is associated with the emergence of Weyl nodes in the electronic structure. 
However, the mechanisms leading to their appearance differ across these temperature regimes. 
In the AFM state, Dirac points first appear due to the breaking of $C_3$ symmetry and are than split into Weyl points through the magnetization process \cite{Soh2018, Su2020}.  

Intriguing is the difference in the sign of these contributions from Weyl nodes in different compounds, for example, they are negative in EuCd$_2$Sb$_2$ (\cite{Su2020} and this work) but positive in EuCd$_2$As$_2$ \cite{Ma2019,Xu2021a}. We suppose that this stems from the fact that the sign of the THE due to the Berry curvature at Weyl points depends on their position relative to the Fermi level, e.g. shown by calculations in \cite{Singh2025}.

\begin{widetext}
    
\begin{figure}[H]
        \centering
		\includegraphics[width=\textwidth]{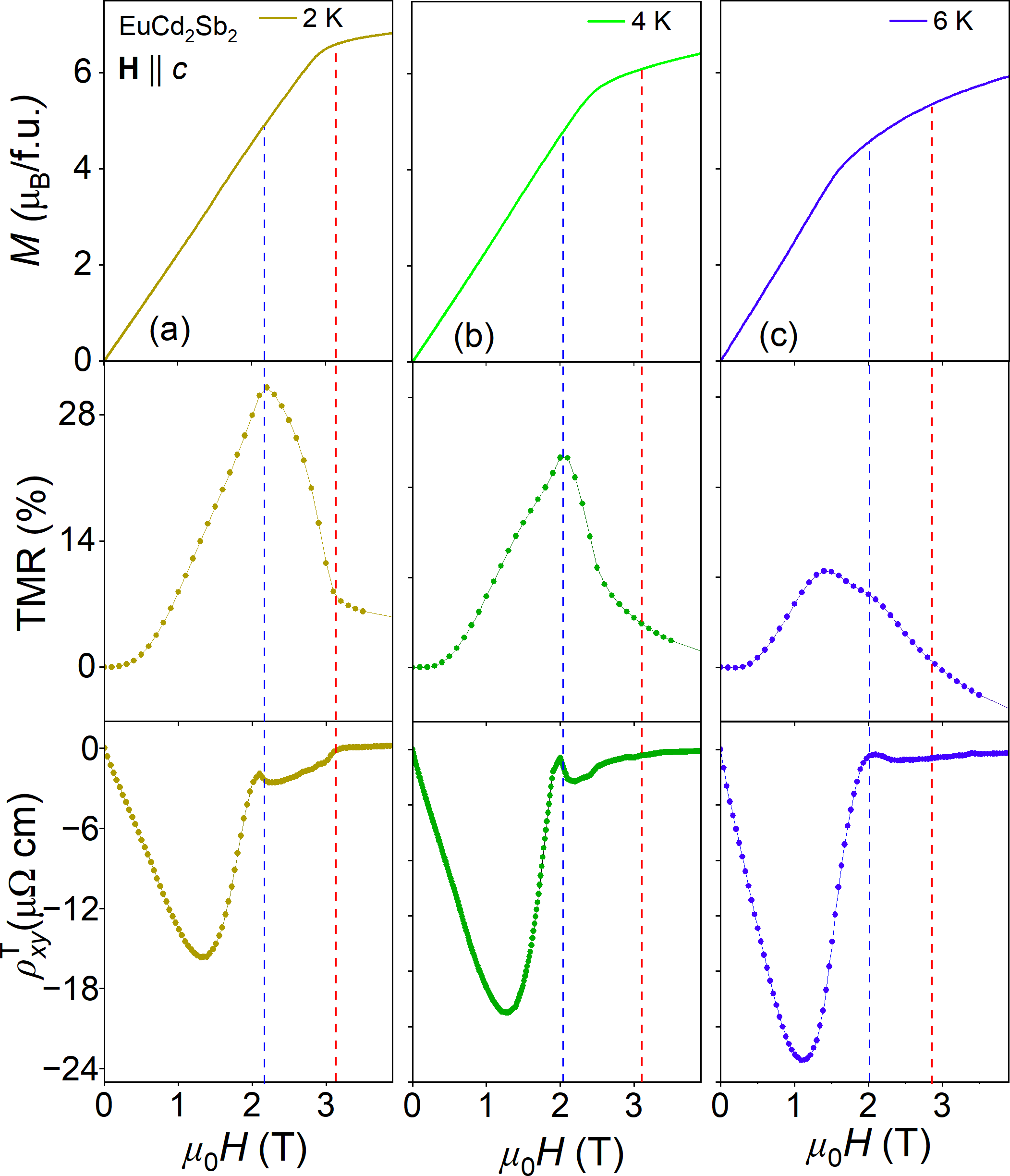}
		\caption{Magnetic field dependence of the magnetization, transverse magnetoresistance and topological Hall resistivity for $T$ = 2\,K (a), 4\,K (b) and  6\,K (c). Red and blue vertical dashed lines mark $H_{\rm s}$ and $H_{\rm s}/\sqrt {2}$, respectively. }
		\label{peak}
\end{figure}

\end{widetext}

Interestingly, at $T<T_{\rm N}$, we observed other small maxima (humps) in $\rho_{xy}^{\rm T}(H)$, indicated with an arrow in Fig.~\ref{TotalHall}(b). 
Similar humps have previously been observed in single crystals and thin films of the same compound \cite{Su2020, Nakamura2024a}, as well as in EuZn$_2$As$_2$ \cite{Yi2023}, EuCd$_2$As$_2$ \cite{Xu2021a}, and EuZn$_2$Sb$_2$ \cite{Singh2024}. They always disappear above $T_{\rm N}$. 
We noticed that these humps have maxima at $H=H_{\rm s}/\sqrt{2}$, similar to those described recently for EuZn$_2$Sb$_2$ \cite{Singh2024}, where they coincided with humps of $\rho_{xx}(H)$ (measured in field transverse to the electric current). Therefore, we plotted $\rho_{xy}^{\rm T}(H)$, TMR$(H)$ and $M(H)$, along each other, for $T<T_{\rm N}$, as shown in Fig.~\ref{peak}.
There is a clearly visible exact alignment of maxima in $\rho_{xy}^{\rm T}$ and TMR with $H_{\rm s}/\sqrt{2}$, for data taken at temperatures of 2 and 4\,K. For $T=6$\,K, $\rho_{xy}^{\rm T}(H)$ has a maximum at $H_{\rm s}/\sqrt{2}$, but TMR displays there a clear, but weaker anomaly (knee). This is most likely because, at the latter temperature, $H_{\rm s}/\sqrt{2}$ is already beyond the range of linearity of $M(H)$. Interestingly, apart from the knee anomaly, TMR recorded at 6\,K also has a pronounced maximum, close to the field, in which $M(H)$ departs from linearity.  

Following the concept devised for EuZn$_2$Sb$_2$ \cite{Singh2024}, where very similar coincidences at $H_{\rm s}/\sqrt{2}$ were observed, we attribute the humps in $\rho_{xy}^{\rm T}(H)$, and the corresponding anomalies in TMR to the scalar spin chirality that emerges within domain walls when an applied field induces local non-coplanar spin textures. The simple calculation shown in Ref.~\cite{Singh2024} demonstrated that such spin chirality is maximal in $H_{\rm s}/\sqrt{2}$ field. On the other hand, it is established that scalar spin chirality provides the Berry phase to the conduction electrons proportional to its magnitude \cite{nagaosa2010anomalous}. 

In the paramagnetic state, Weyl nodes may be induced by spin fluctuations, this scenario has recently been supported by combined DFT calculations and ARPES measurements for both EuCd$_2$Sb$_2$ and EuCd$_2$As$_2$ \cite{Ma2019}, and seems suitable for other similar compounds, including EuZn$_2$Sb$_2$. 
To estimate the upper temperature limit for that effect we analyzed the derivative, $d\chi^{-1}(T)/dT$ and found that below certain temperature $T_{\rm sf}\approx$40\,K it abruptly drops from its constant value, marking a significant deviation from the linear behavior of $\chi^{-1}(T)$. The $T_{\rm sf}$ is also the temperature when the upturn in $\rho_{xx}(T)$ starts (cf. Fig.~%\ref{Hall-decompo}
S3(c) in the Supplemental Material \cite{SupM}). It can thus be interpreted as the point, above which the spin fluctuations' effect on both magnetization and resistivity becomes negligible.   
The effect of spin fluctuations on resistivity can be roughly estimated with $\Delta\rho_{sf}=[{\rho(T_{\rm N})}/ {\rho(T_\text{sf})}]-1$. For EuCd$_2$Sb$_2$ it is $\Delta \rho_{sf}=0.044$, slightly larger than 0.016 estimated for EuZn$_2$Sb$_2$ \cite{Singh2024} (which also exhibited sizable THE in the paramagnetic state), but much smaller than 23 observed for EuZn$_2$As$_2$ and 1 for EuCd$_2$As$_2$. 
\section{Summary and conclusion}\label{Summary}
In this study we grew and characterized the single crystals of EuCd$_2$Sb$_2$. The heat capacity we measured reconfirmed the magnetic ordering below $T_{\rm N}=7.4$\,K and revealed maximum magnetic entropy near the ideal value \textit{R\,}ln(8) for Eu$^{+2}$ ion. We also found that acoustic phonon modes are dominant at high-temperatures. The following magnetization measurements reconfirmed the AFM ground state with localized moments of Eu$^{+2}$ ions and showed how magnetic structure changes in applied magnetic field. 
Next, we investigated the electrical resistivity in longitudinal and transverse fields, as well as Hall resistivity. Enhanced negative magnetoresistance in the longitudinal field corroborated the occurrence of Weyl nodes. The topological component was extracted from total Hall resistivity. Prominent anomalies were found in field variations of TMR, LMR and $\rho_{xy}^{\rm T}$ in the applied field range where the magnetic structure undergoes strong modification. 
After analyzing the behavior of the extrema of these anomalies we propose that at least three intrinsic mechanisms contribute to the THE in EuCd$_2$Sb$_2$. In the AFM state they are the Berry curvature due to Weyl nodes originating from broken $C_{3}$ symmetry and the Berry curvature emerging from the scalar spin chirality within domain walls. Above $T_{\rm N}$ it is the Berry curvature due to Weyl nodes induced by the spin fluctuations.
\\\\\textbf{DATA AVAILABILITY}\\
The data that support the findings presented above are openly available \cite{DATASET}.\\\\
\textbf{ACKNOWLEDGMENT}\\
This study was supported by the National Science Centre (Poland) under grant 2021/41/B/ST3/01141.

\newpage
\clearpage
\onecolumngrid
%    \begin{widetext}
 \begin{center}
  \textbf{\large Supplemental Material for}\label{Supplemat}\\[.4cm]
  \textbf{\large{Origin of the large topological Hall effect in the EuCd$_2$Sb$_2$ antiferromagnet}}\\[.2cm]
Faheem Gul, Orest Pavlosiuk, Tetiana Romanova, Dariusz Kaczorowski and Piotr Wiśniewski\\
  {\itshape Institute of Low Temperature and Structure Research, Polish Academy of Sciences, Wrocław, Poland\\
  }
(\today)
\\[1cm]
\end{center}

\setcounter{figure}{0}
\renewcommand{\thefigure}{S\arabic{figure}}
\setcounter{section}{0}
\renewcommand{\thesection}{S\arabic{section}}
\setcounter{table}{0}
\renewcommand{\thetable}{S\arabic{table}}
\setcounter{page}{1}
\renewcommand{\thepage}{s\arabic{page}}

%%%%%%%%%%%%%%%%%%%%%%%%%%%%%%%%%%%
%%%%%%%%%%%%%%%%%%%%%%%%%%%%%%%%%%%
%%%%%%%%%%%%%%%%%%%%%%%%%%%%%%%%%%%

\section{Crystal Growth and Characterization}\label{CGnC_SupM}
Single crystals of EuCd$_2$Sb$_2$ were grown from the Sn-flux. Eu, Cd, Sb, and Sn were taken in 1:2:2:4 molar ratios, put inside an alumina crucible and sealed in evacuated quartz ampule. The ampule was heated at 20\,°C/h rate up to 1000\,°C, kept at that temperature for 24 hours, and then cooled to 400\,°C at 2\,°C/h. The flux was removed by centrifugation. Fig.~S1 shows elemental composition (a) and Laue diffraction pattern (b). Trigonal crystal structure (with $P\bar{3}m1$ space group) was reconfirmed by X-ray diffraction on a single crystal, using Mo-$K\alpha$ radiation. The obtained lattice parameters are shown, and compared to those previously reported in Table S1.  

\begin{table}[!h]
\begin{center}
\caption{Lattice parameters obtained from the refinement on 52 reflections collected with single-crystal x-ray diffractometer, and compared to those previously reported. The quality of our refinement (UB fit) was 98.08\%.}

\begin{tabular}{l l l c} 
\textit{a} (\AA)  & \textit{c} (\AA)& cell volume (\AA$^{3}$)& Ref. \\ 
\hline\hline
4.683(10) & 7.766(11) & 147.5(4) & this work  \\ 
4.698(1) & 7.723(1) &147.922(66) &\cite{Artmann1996}  \\
4.699(2)  &  7.725(2) & 147.72(13)&\cite{Schellenberg2011}  \\
4.7030(9) & 7.7267(18) & 148.004(66)& \cite{Soh2018}  \\
4.6926(6) & 7.7072(8) & 147.013 (40) & \cite{Su2020} \\  
\hline
\end{tabular}
\end{center} \label{TableS1}
\end{table}

\begin{figure}[H]
 \centering
		\includegraphics[width=0.85\textwidth]{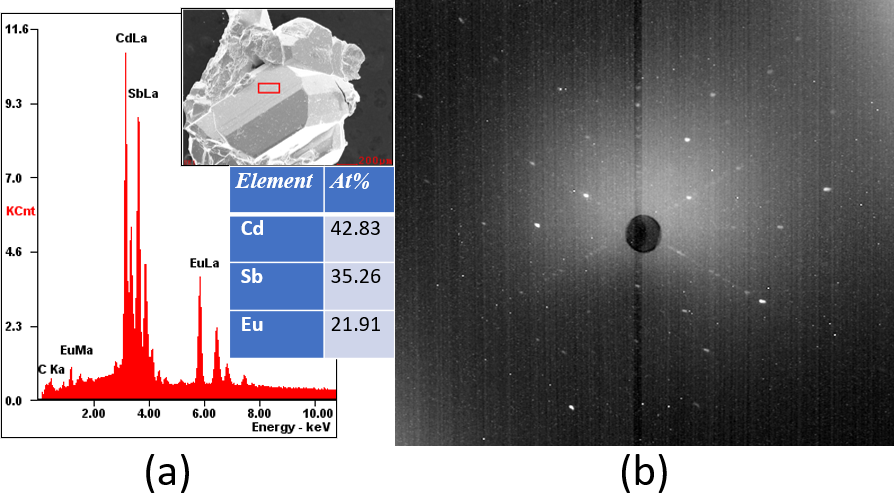}
			\caption {(a) Analysis of an EDX spectrum collected on the EuCd$_2$Sb$_2$ single crystal pictured in inset (the red rectangle indicates the area of measurement) yields the elemental composition close to 1:2:2 (given in the table, errors are about 2\%). (b) Laue diffraction pattern, taken with an incident X-ray beam close to [001] crystallographic direction.} 
			%\label{FigureS1}
\end{figure}

\section{Magnetic Properties}\label{MP_SupM}
Figure~\ref{MH-MT} in the main text shows temperature dependence of the inverse magnetic susceptibility, $\chi^{-1}(T)$, magnetic susceptibility, $\chi(T)$ and magnetization, $M(H)$. The Curie-Weiss law $\chi^{-1}= (T-\theta_{\rm P})/C$, was fitted in the region from 40 to 300\,K to data taken in the magnetic field $\mu_0H$=0.5\,T applied perpendicular to the \textit{c}-axis, as shown Fig. \ref{MH-MT}(a), where $C\,(=\mu_0 N\mu_{\rm eff}^{2}/3k_{\rm B})$ is Curie parameter, and $\theta_{\rm P}$ is the paramagnetic Curie-Weiss temperature. Least squares fitting yielded $\theta_{\rm P}$= 3.45\,K and $C$ = 8.08\,$\mu^{2}_{\rm B}$, both similar to previously reported \cite{Soh2018, Su2020} (note that the sign of $\theta_{\rm P}$ was mistakenly inverted in ref.~\cite{Soh2018}). Positive value of $\theta_{\rm P}$ shows the dominant role of ferromagnetic interactions.
The experimentally derived $\mu_{\rm eff} =8.04\,\mu_{\rm B}$ is close to $7.94\,\mu_{\rm B}$, the theoretical value of Eu$^{+2}$ ion calculated as $\mu_{\rm eff}=g\sqrt{S(S+1)}\mu_{\rm B}$ with $g=$2 and $S=$7/2.
 The temperature-dependent susceptibility $\chi(T)$ measured in $\mu_0H=0.05$\,T confirms the AFM ordering at 7.4\,K as illustrated in Fig.~1(a) in the main text, in accordance with the previous reports \cite{Su2020, Ohno2022, Soh2018}. The ordering temperature decreases with increasing magnetic field for its both directions, see Fig.~\ref{MH-MT}(b) and inset of Fig.~\ref{MH-MT}(c) in the main text. The resulting \textit{M-T} phase diagram for \textbf{H} $\parallel$ \textit{c}-axis, is shown in Fig.~\ref{origin-signature}(c), in the main text.

The field-dependent magnetization, $M(H)$, shows no saturation in fields up to 7\,T. However, there is clearly visible linear trend towards the saturation in $\mu_0H_{\rm s}=1.9$\,T for  $\mathbf{H}\!\perp$\textit{c}-axis (shown in of Fig.~\ref{MH-MT}(c)), and in $\mu_0 H=3.1$\,T for $\mathbf{H}\!\parallel$\textit{c}-axis at 2\,K (shown in Fig.~\ref{MH-MT}(d)). Figure~S2 shows how we estimated the saturation fields. For 2, 4 and 6\,K, the saturation is estimated in $\mu_0H_{\rm s}=3.15$, 3.05 and 2.95\,T, respectively, when $\mathbf{H}\!\parallel\!c$-axis. The linear magnetization in the low  magnetic field region and $H_{\rm s}$ decreasing with increasing temperature are typical features of the AFM phase. %\newpage
\begin{figure}[h]
\centering
\includegraphics[width=0.6\textwidth]{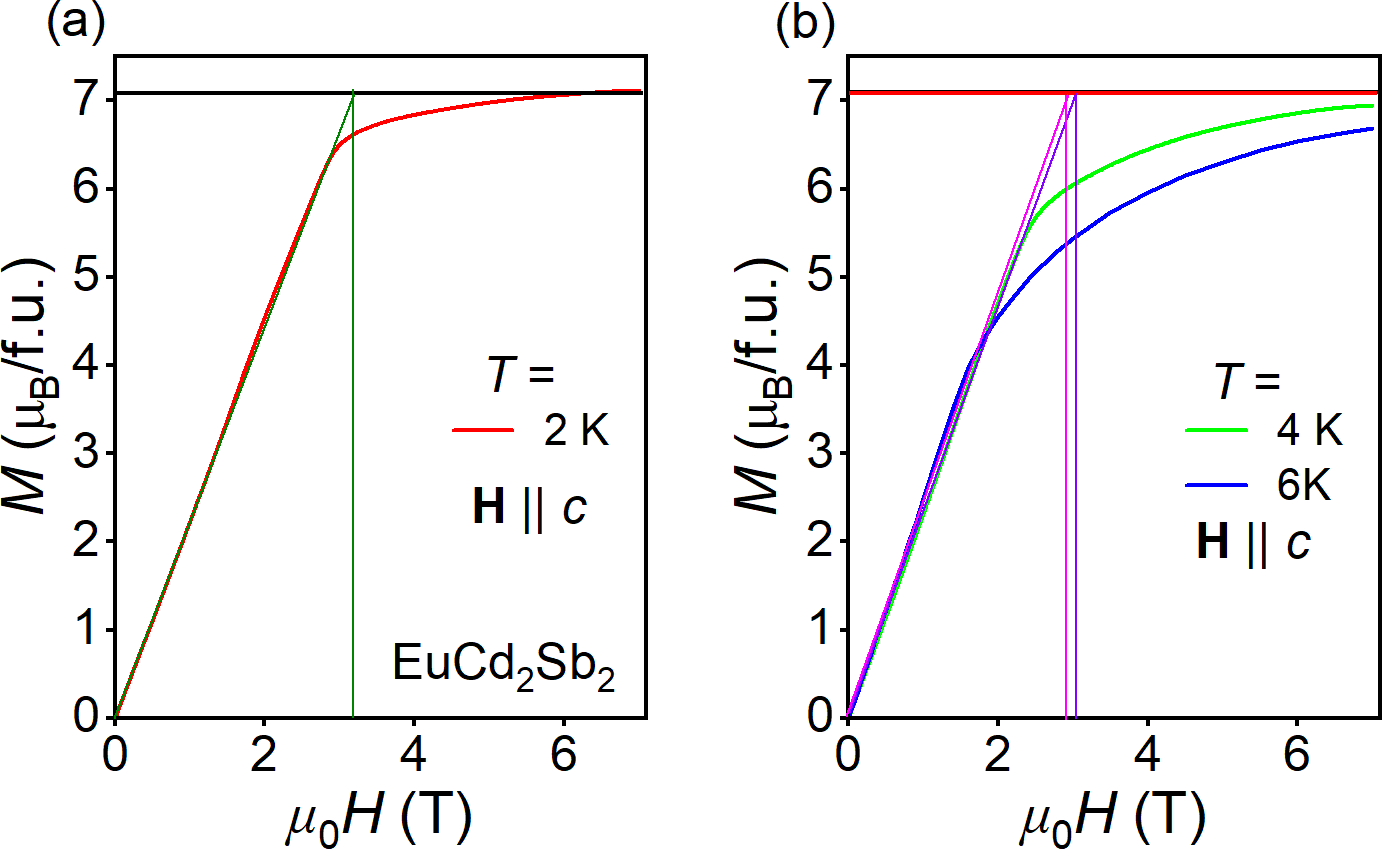}
\caption{The estimation of $H_{\rm s}$: for 2\,K  (a), 4\,K (violet) and 6\,K (pink) (b)}
%\label{FigureS2}
\end{figure} 
\section{Specific heat}\label{specific heat SM}
The characteristic parameters obtained from the Debye-Einstein model for Eu-based 122 compound are compared in Table S2. Pakhira et el. \cite{Pakhira2022} employed two different Debye-Einstein models, one with single Einstein term and the other with two Einstein terms for EuMg$_2$Sb$_2$, which is a semiconductor, thus $\gamma=0$.
\begin{table}[H]
\begin{center}
\caption{Characteristic parameters from the specific heat analysis for Eu-based 1:2:2 compounds.}
\label{TableS2}
\begin{tabular}{c c c c c c }
%\hline
 &$\theta_{\rm E}$ (K) & $\theta_{\rm D}$ (K) & $\gamma$\,(mJ\,mol$^{-1}$K$^{-2}$) & $p$ & Ref.\\ [1ex]
 \hline
 \hline
 EuCd$_2$Sb$_2$ & 81 & 228 & 8.1 & 0.36 & this work \\ [0.5ex]
 %\hline
 EuZn$_2$Sb$_2$ & 77 & 248 & 4.3 & 0.2 &\cite{Singh2024}\\[0.5ex]
 %\hline
 EuMg$_2$Sb$_2$ & 111 & 415  & 0 & 0.48  &\cite{Pakhira2022} \\
  &~~~~ 94, 749 ~~~~& 309  & 0 & 0.3, 0.08  & \cite{Pakhira2022}\\[0.5ex]
 \hline\hline
\end{tabular}
\end{center}
\end{table}

\section{Decomposition of the Hall resistivity}\label{separation}
The Hall resistivity $\rho_{xy}(H)$ was measured in magnetic field up to $\mu_0H=9$\,T, and is the sum: $\rho_{xy}=\rho^{\rm o}_{xy}+\rho_{xy}^{\text{A}}$. Here $ \rho^{\rm o}_{xy }=\mu_0R_{\rm H}H$ is the ordinary Hall resistivity, with $R_{\rm H}$ being the ordinary Hall coefficient. $\rho^{\rm A}_{xy}$ is the anomalous Hall resistivity consisting of two terms:  $\rho_{xy}^{\text{A}}= S_{\rm H }\rho_{xx}^{2}M +\rho_{xy}^{\rm T}$, where $S_{\rm H }$ is the conventional anomalous Hall coefficient, $M$ is the magnetization, and $\rho_{xy}^{\rm T}$ is the topological Hall resistivity. 
The slope of linear $\rho_{xy}(H)$ from high magnetic field region was used to calculate $R_{\rm H}$.
The obtained $R_{\rm H}$ served to subtract $\rho^{\rm o}_{xy}$ from $\rho_{xy}$. The ranges of the magnetic field used for linear fitting to $\rho_{xy}(H)$ are given for various temperatures in the Table~\ref{regions} of the main text. 
Next, to find the $S_{\rm H}$, we used the relation $\rho_{xy}^{\rm A}= {S_{\rm H }}\rho_{xx}^{2}M+\rho_{xy}^{\rm T}$, and assumed that $\rho_{xy}^{\rm T}$ vanishes in  magnetic field above $H_{s}$. This assumption stands on the basis that the linear slope of the $\rho_{\rm xy}$ is almost temperature independent, therefore, above $H_{s}$, $\rho_{xy}^{\rm A} =\rho_{\rm xy}^{\rm CA}= {S_{\rm H }}\rho_{xx}^{2} M$. The value of  $\rho_{xy}^{\rm A}$  at the highest magnetic field was used to estimate $S_{\rm H}$. Then the $M(H)$ and $\rho_{xx}(H)$ data were used to calculate $\rho_{xy}^{\rm A}$.  Each step of the extraction of $\rho_{\rm xy}^{\rm T}$ for $T$ = 2\,K  is shown in Fig.~S3(a). The same $S_{\rm H}$ was used to subtract $\rho_{\rm xy}^{\rm CA}$ from $\rho_{\rm xy}^{\rm A}$ at higher temperatures.

\begin{figure}[]
	\centering
\includegraphics[width=0.8\textwidth]{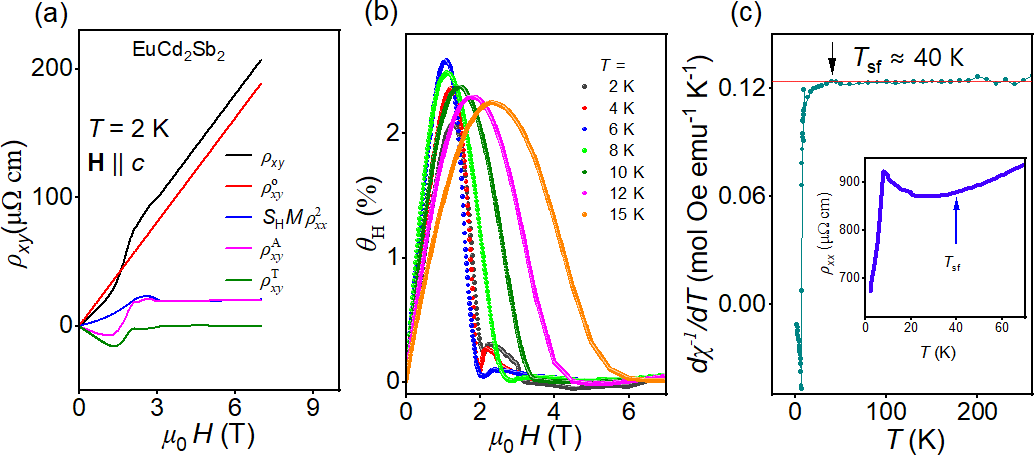}
   \caption{(a) Components of Hall resistivity taken into account for the extraction of $\rho_{xy}^{\rm T}(H)$ for $T$=2\,K. (b) Magnetic field dependence of the Hall angle ($\theta_{\rm H}$) (c) The first-order derivative of the inverse magnetic susceptibility. The inset shows the temperature dependence of the longitudinal resistivity ($\rho_{\rm xx}$).} 
   %\label{FigureS3}
\end{figure}

%\FG{\section{Magnetoresistance} Magnetoresistance (MR\%) for $\mathbf{H}\perp c$ and $\perp \mathbf{j}$ electrical configuration is illustrated in Fig.~\ref{MR_supM}. The anomaly in MR falls at $\sqrt{2}$ position of the field, where the peak diminishes, for this configuration, similar to two other configurations shown in the main text. However, the fourth possible configuration i.e., $\mathbf{H}\parallel c$ and $\parallel \mathbf{j}$ can be realized using modern techniques, to consolidate this proposal further.}
\section{Role of spin fluctuations} 
The derivative of the inverse of magnetic susceptibility ($d\chi^{-1}/dT$) is shown in Fig.~S3(c). The black arrow marks the temperature, $T_{\rm sf}$, where the $\chi^{-1}$ visibly departs from linearity. The inset shows the temperature dependence of the longitudinal resistivity, $\rho_{\rm xx}(T)$, at $T\leq70$ K. The blue arrow indicates the start of the $\rho_{\rm xx}$ upturn upon lowering temperature at the same $T_{\rm sf}$. Both phenomena are due to spin fluctuations.
%\begin{figure}[]
%	\centering \includegraphics[width=0.35\textwidth]{MR_B_perp_c_j.png}
%   \caption{Field dependence of MR$(H)$ for $\mathbf{H}\perp c$ and $\perp \mathbf{j}$.} \label{MR_supM}
%\end{figure}
%
\section{Conversion of resistivity to conductivity}
Figure~S4 shows the field dependence of the $\rho_{\rm xx}$ in longitudinal (a) and transverse fields (c), and corresponding conductivities, (b) and (d), respectively, for various temperatures from 2 to 150 K. Magnetoconductivity shown in  Fig.~S4(b) and (d), was calculated as: $\sigma_{xx}= \rho_{xx}/ (\rho_{xy}^{2}+\rho_{xx}^{2})$. Figure~S5(a) shows the temperature dependence of the longitudinal resistivity $\rho_{xx}$ and corresponding conductivity (calculated as $\sigma_{xx}=1/ \rho_{xx}$). The residual resistivity ratio ($\rho_{xx}(300\rm K)/\rho_{xx}(40\rm K)$) for our sample is 1.93, very close to 2 reported in \cite{Su2020}, indicating a very similar crystal quality. 
 
Field-dependent anomalous Hall conductivity, total Hall conductivity, and topological Hall conductivity for various temperatures, calculated as $\sigma_{xy}^{\rm A}= -\rho_{xy}^{\rm A}/ ({\rho_{xy}^{\rm A}}^{2}+\rho_{xx}^{2})$, $\sigma_{xy}= -\rho_{xy}/ (\rho_{xy}^{2}+\rho_{xx}^{2})$, and $\sigma_{xy}^{\rm T}= -\rho_{xy}^{\rm T}/({\rho_{xy}^{\rm T}}^{2}+\rho_{xx}^{2})$, 
are shown in Fig.~S5(b), inset to the S5(b) and S5(c), respectively.  

\begin{figure}[h]
	\centering
	%\subfigure{
		\includegraphics[width=0.5\textwidth]{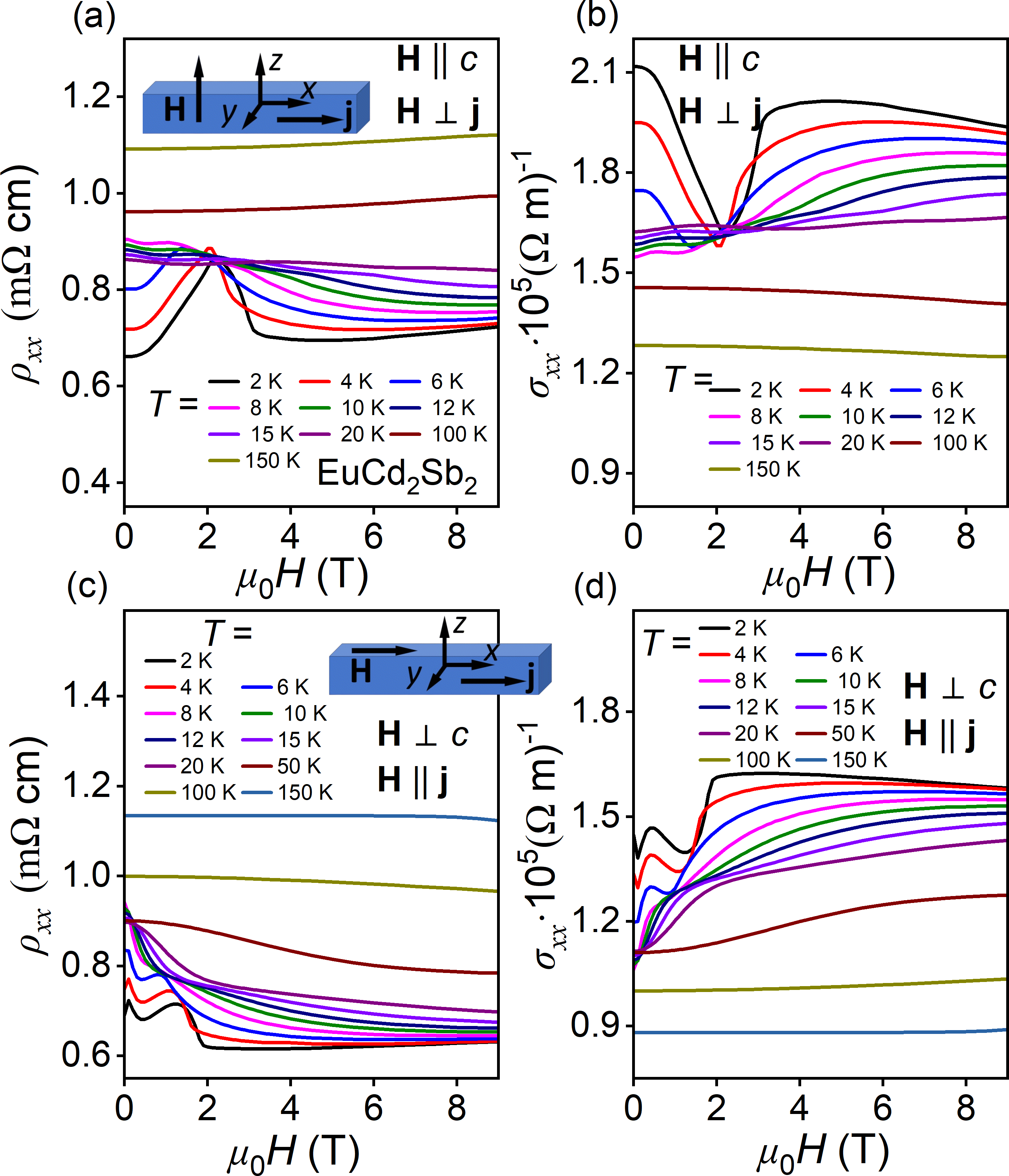}
	\caption{Longitudinal and transverse magnetoresistivity and magnetoconductivity as a function of magnetic field for various temperatures.  }
	%\label{FigureS4}
\end{figure}
\begin{figure}[h]
	\centering
	%\subfigure{
		\includegraphics[width=\textwidth]{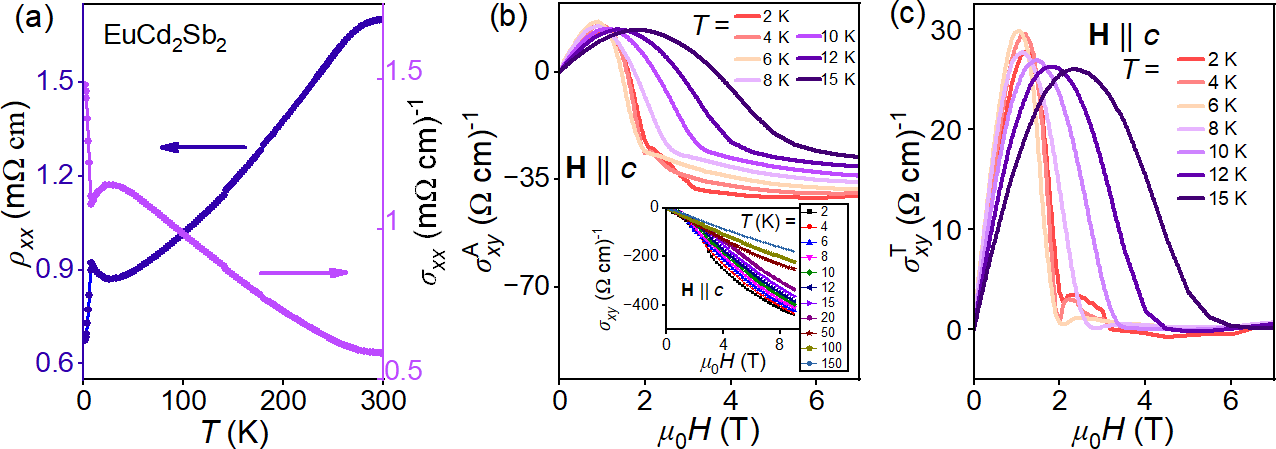}
	\caption{ (a) Temperature dependence of the resistivity and the corresponding conductivity, from 2 to 300\,K. (b) and (c) The magnetic field-dependence of the anomalous Hall (inset: shows magnetic field dependence of the total Hall conductivity) and topological Hall conductivities, respectively, for various temperatures.}
	%\label{FigureS5}
\end{figure}
\end{document}